\newcommand{\be}{\begin{equation}}\newcommand{\ee}{\end{equation}}
\newcommand{\bea}{\begin{eqnarray}}\newcommand{\eea}{\end{eqnarray}}
\newcommand{\nn}{\nonumber}\newcommand{\p}[1]{(\ref{#1})}
\documentstyle[12pt]{article}
\begin{document}

% SIDE MARGINS :

\if@twoside
   \oddsidemargin  0.5 cm
   \evensidemargin 0cm
\else
   \oddsidemargin  0.5cm
   \evensidemargin 0.5cm
\fi

% VERTICAL SPACING :

\topmargin  0truecm
\headheight 0pt
\headsep    0pt
\topskip    1pt

% DIMENSION OF TEXT :

\textheight 680pt
\textwidth  16.5cm

\begin{titlepage}
  \begin{center}
    \font\GIANT=cmr17 scaled\magstep4
	{\GIANT U\kern0.8mm N\kern0.8mm I\kern0.8mm V\kern0.8mm %
	E\kern0.8mm R\kern0.8mm S\kern0.8mm I\kern0.8mm %
	T\kern0.8mm %
	\setbox0=\hbox{A}\setbox1=\hbox{.}%
	\dimen0=\ht0 \advance \dimen0 by -\ht1%
	\makebox[0mm][l]%
	{\raisebox{\dimen0}{.\kern 0.3\wd0 .}}A\kern0.8mm
	T\kern5mm
	B\kern0.8mm O\kern0.8mm N\kern0.8mm N\kern0.8mm
	} \\[7mm]
    {\GIANT
    	P\kern0.8mm h\kern0.8mm y\kern0.8mm s\kern0.8mm i\kern0.8mm
	k\kern0.8mm a\kern0.8mm l\kern0.8mm i\kern0.8mm s\kern0.8mm
	c\kern0.8mm h\kern0.8mm e\kern0.8mm s\kern5mm
	I\kern0.8mm n\kern0.8mm s\kern0.8mm t\kern0.8mm
	i\kern0.8mm t\kern0.8mm u\kern0.8mm t\kern0.8mm
	} \\[1.5cm]
    {\Large \bf
Chiral fermion action  \\[0.5mm]		%<---
 with (8,0) worldsheet supersymmetry}\\[9mm]	%<---
{\bf E. Ivanov${}^{(a)}$
and E. Sokatchev${}^{(b)}$} \\[5mm]

{ ${}^{(a)}$\it Laboratory of Theoretical Physics, Joint Institute for
Nuclear Research, 141980 Dubna near Moscow, Russia}\\[1mm]

{${}^{(b)}$\it Physikalisches Institut, Universit\"at Bonn, Nussallee 12,
D-53115 Bonn, Germany}\\[5mm]

{\bf Abstract}  \\[1.5mm]%<---
  \end{center}
{We propose an action describing chiral fermions with an arbitrary gauge
group and with manifest $(8,0)$ worldsheet supersymmetry. The form of the
action is inspired by and adapted for completing the twistor-like formulation
of the $D=10$ heterotic superstring.}

  \begin{figure}[b]
    \begin{minipage}{\textwidth}
	\begin{raggedright}
	\begin{tabular}{@{}l@{}}
		Post address:  \\
	        Nussallee 12   \\
	        D-53115 Bonn   \\
	        Germany      \\
	\end{tabular}
	\end{raggedright}
	\hfill
	\parbox{5.5cm}
	{
	    \vbox to 2cm
	    {
    		\vskip -2truecm
		\hbox to 5.5cm
		{
%		    \special {psfile=is.ps voffset=-155}
%                    \psfig{figure=tex_inputs:is.ps}
		    \hfill
		}
		\vfill
		\centering
	    }
	}
	\hfill
	\begin{raggedleft}
        \begin{tabular}{@{}l@{}}
	    BONN-TH-94-10         \\
	   %	    BONN-ME-xx-xx	     	    \\		%<---
	    Bonn University          	    \\
	    June 1994                      \\		%<---
%	    {\footnotesize ISSN-0172-8741}  \\		%<--- f"ur BONN-IR
%	    {\footnotesize ISSN-0172-8733}  \\		%<--- f"ur BONN-HE
%	    {\footnotesize ISSN-0936-2797}  \\		%<--- f"ur BONN-ME
	\end{tabular}
	\end{raggedleft}
    \end{minipage}
  \end{figure}
\end{titlepage}

\newpage
\setcounter{page}{1}

\noindent {\bf 1. Introduction.} Chiral fermions are a necessary
ingredient of the self-consistent
quantum heterotic superstring theory. They serve to compensate the
conformal anomaly and ensure that the critical dimension of space-time
is 10 \cite{kniga}.

Recently various  twistor-like formulations of the $D=10$
heterotic string at the classical level have been proposed
\cite{Ton,DGHS,gr}.  The basic
attractive feature of twistor-like formulations of the heterotic superstring
\footnote{Recently several attempts have been made to adapt the twistor-like
approach to non-heterotic superstrings \cite{nonhet} or $p$-branes
\cite{p-b}, but in our opinion it is still not entirely clear to what
extent these formulations are self-consistent.} is the
trading of the local $\kappa$-symmetry of the $D=10$ superstring action
for $N=(8,0)$ local supersymmetry of the worldsheet. The twistor-like
mechanism was first discovered by Sorokin et al \cite{volk}
in the context of a $D=3,4$ superparticle with local $N=1,2$ worldline
supersymmetry and then generalized to the $D=6$ \cite{DS} and $D=10$
\cite{GS} superparticle with $N=4$ and $N=8$ worldline supersymmetry and
to heterotic superstrings in $D=4$ \cite{IK} and $D=6$ \cite{DIS} with
$N=2,4$ worldsheet supersymmetry (see also \cite{berko}). It
unveiled for the first time the transparent geometric meaning of
$\kappa$-symmetry.  The fact that all symmetries of the
superstring action become manifest and geometrically interpretable
within a twistor-like formulation gives us hope that the latter
is most appropriate for covariant quantization of the heterotic
superstring.

Keeping in mind this ambitious though still remote prospect, one may wonder
how to consistently incorporate, to begin with at the
classical level, the chiral fermions into the twistor-like
formulation of the heterotic
superstring while preserving all symmetries and geometric features of the
latter. An early attempt in this direction was made by Tonin \cite{Ton}.
There the chiral fermion superfield equation of motion was introduced
into the action with a Lagrange multiplier. Unfortunately, the latter
turned out to propagate unwanted degrees of freedom.
Another approach to this problem was proposed
by Sorokin and Tonin \cite{ST}. This time the field equations of the chiral
fermions are derived from a quadratic $(8,0)$ superfield constraint, once
again added to the superstring action with a Lagrange
multiplier. To make the derivation unambiguous and to avoid the
propagation of extra degrees of freedom, they have to restrict themselves
only to supersymmetric solutions of their constraint and to resort
to a rather subtle positiveness argument.

A different $(8,0)$ action for chiral
fermions has recently been proposed by Howe \cite{Howe}. It is a standard
bilinear superfield action with a specially
chosen operator made out of up to 7 spinor derivatives. The action has a
large abelian gauge invariance which helps to restrict the on-shell
content of the theory to just chiral fermions. The superfields incorporating
these fermions must satisfy a certain constraint which effectively ties up
the internal symmetry index on the fermion fields to the (local) $SO(8)$
symmetry group of the worldsheet. As remarked in \cite{Howe}, this results
in a severe restriction on
the internal symmetry group (it should not be bigger than $SO(4)$).

In the present letter we propose another chiral fermion action with local
$(8,0)$ worldsheet supersymmetry and with an arbitrary gauge group.
It is a  modification of the one of Sorokin and Tonin \cite{ST}.
Like in \cite{ST}, an appropriate
superfield terms is included into the Wess-Zumino term of the twistor
superstring action sharing the same Lagrange multiplier superfield.
As shown in \cite{DGHS}, after fixing a certain gauge this Lagrange
multiplier is reduced just to the
string coupling constant. Then the new term produces  the
standard kinetic term for chiral fermions as well as an auxiliary field term.
We stress that the chiral fermion modification
of the superstring Wess-Zumino term we propose is essentially different
from that in \cite{ST}. It is chosen so that requiring compatibility
with the standard constraints on the background supergravity three-form
{\it does not} imply  equations of motion for the chiral
fermions (as it does in \cite{ST}). Instead, the
integrability condition results in a
quadratic constraint on the
chiral fermions superfields which simply reduces the off-shell
component content of the
superfields (quite similarly to standard linear irreducibility
constraints). This constraint turns out to be also necessary for the
off-shell local worldsheet supersymmetry of the chiral fermion action.
The realization of off-shell supersymmetry in this action has some unusual
features which we discuss in detail.

\vspace{0.4cm}
\noindent {\bf 2. Preliminaries.} We start by briefly recalling the
structure of the heterotic
superstring action in the twistor-like formulation as it has been
given in \cite{DGHS}. The action consists of three parts, each of them given
by an integral over the $(8,0)$ worldsheet superspace
\begin{equation} \label{twistact}
S = S_1 + S_2 + S_{wz} \;.
\ee
For our purposes it will be important to know
the explicit form of $S_{wz}$
\begin{equation} \label{wz}
S_{wz} = \int d^2 x d^8 \theta P^{MN} [\; B_{MN} +
E_{[M}^{\;\;+} E_{N \}}^{\;\;-} {\mbox
e}^\Phi E^{\underline{a}}_- E_{+\;\underline{a}}
- \partial_{[M}
Q_{N\}}\;]\;.
\end{equation}
Here and in what follows non-underlined indices refer to the worldsheet
superspace $\{ z^M \} = \{ x^-, x^+, \theta^\mu \}$, $\mu =1,...,8 $, i.e.
$M = -,+,\mu$, while underlined ones refer to the
target $D=10$ superspace $z^{\underline{M}}$. For instance,
$\underline{a} =0,1,...,9$ is the vector tangent space index. The symbol
$[MN\}$ means graded antisymmetrization. The quantities $B_{MN}$ and
$E_{\pm}^{\underline{a}}$ are, respectively,
the worldsheet superspace pull-backs of the background $D=10$ supergravity
two-form $B_{\underline{MN}}$ and vielbeins $E_{\underline{M}}^{\;\;
\underline{A}}$
\begin{equation}
B_{MN} = (-)^{M (N + \underline{N})} \partial_N z^{\underline{N}}
\partial_M z^{\underline{M}} B_{\underline{MN}} \;,  \;\;\;
E_{\pm}^{\underline{A}} = E_\pm^M\partial_M z^{\underline{N}}
E_{\underline{N}}^{\underline{A}}\;,
\end{equation}
$E_{M}^{\;\; \pm}$ are elements of the vielbein and $E_\pm^M$ of the
inverse vielbein
matrix on the worldsheet, $\Phi$ is the dilaton field of
background supergravity, $Q_N$ and $P^{MN}$ are worldsheet
superfield Lagrange multipliers, whose r\^ole will be explained later on.

The term $S_1$ in (\ref{twistact}) produces the so called geometro-dynamical
constraint
\begin{equation}\label{gd}
E_\alpha^{\underline{a}} \equiv E_\alpha^M \partial_M z^{\underline{N}}
E_{\underline{N}}^{\underline{a}} = 0
\end{equation}
on the spinor pull-back of the vector target superspace vielbeins. It has as
a corollary a twistor representation for the left-handed Virasoro vector
\begin{equation}\label{twlk}
E_-^{\underline{a}} = {1\over 8} E_\alpha^{\underline \alpha}
\gamma^{\underline{a}}_{\underline{\alpha\beta}} E_\alpha^{\underline\beta}
\end{equation}
which provides a {\it twistor-like solution} to the left-handed
Virasoro constraint $E_-^{\underline{a}} E_{-\underline{a}} = 0$ of the
superstring. The term $S_2$ in (\ref{twistact})
enforces an irreducibility
condition on the spinor component $e_{\alpha}^+$ of the worldsheet inverse
vielbein (see eq. (\ref{cnstr}) below); the latter covariantizes
the worldsheet superspace derivatives with respect to arbitrary shifts
of the right-handed coordinate $x^+$ and thus provides the Lagrange
multiplier for the right-handed
Virasoro constraint. More details on these two terms can
be found in \cite{DGHS}.

The term (\ref{wz}) is the leading one in the twistor superstring action,
since it produces the entire superstring
component action. It is instructive to discuss in some detail how this works,
because we will use the same mechanism for generating the
chiral fermion component action.

The structure of $S_{wz}$ is entirely specified by the requirement
that the result of varying with respect to the Lagrange
multiplier $P^{MN}$,
\begin{equation} \label{bequat}
B_{MN} = \partial_{[M} Q_{N \}} -
E_{[M}^{\;\;+} E_{N \}}^{\;\;-} {\mbox
e}^\Phi E^{\underline{a}}_- E_{+\;\underline{a}}\;.
\end{equation}
is consistent with the $D=10$ supergravity
constraints on the three-form
field strength $H_{\underline{KMN}}$ of the two-form $B_{\underline{MN}}$
\cite{10sg}. This means that eq. (\ref{bequat}) implies the integrability
condition
\begin{equation} \label{hequat}
H_{KMN} \equiv  \partial_{[K} B_{MN \}} =
- \partial_{[K} (E_{M}^{\;\;+} E_{N \}}^{\;\;-} {\mbox
e}^\Phi E^{\underline{a}}_- E_{+\;\underline{a}})\;.
\ee
Using the geometro-dynamical constraint (\ref{gd}) and its corollary
(\ref{twlk}), one
can show that equation (\ref{hequat}) is consistent with the
$D=10$ background supergravity constraints and is actually equivalent to them.
When  (\ref{hequat}) holds the action term $S_{wz}$ is invariant
under the following gauge transformation
\begin{equation} \label{gauge1}
\delta P^{MN} = \partial_K \Sigma^{MNK}\;,
\ee
where  $\Sigma^{KMN}(z)$ is a totally (graded) antisymmetric
superfield.

To see how (\ref{wz}) produces the standard component heterotic
superstring action, one varies with respect to the Lagrange multiplier
$Q_{M}$ and obtains
\begin{equation}
\partial_N P^{MN} = 0\;.
\end{equation}
The general solution of this equation is given by
\begin{equation} \label{cohom1}
P^{MN} =  \partial_K \Lambda^{MNK} + \theta^8 \delta_+^{\;\;[M}
\delta_-^{\;\;N \}} T \;.
\end{equation}
The second (cohomological) term contains the constant $T$,
\begin{equation}
\partial_+ T = \partial_- T = 0\;.
\end{equation}
Then one substitutes this solution back into $S_{wz}$ and
observes that the first term in (\ref{cohom1}) can be completely
gauged away using the gauge freedom (\ref{gauge1}). As a result
in this gauge in $P^{MN}$ there survives only one component
\begin{equation} \label{pgauge}
P^{+-} = \theta^8 T
\end{equation}
and $S_{wz}$ (\ref{wz}) is reduced to the ordinary $x$-space integral
of the first component of the superfield expression within the
square brackets. It contains, in particular, the superstring WZ term
\begin{equation}
S_{wz} = T \int d^2 x \;\partial_- z^{\underline{N}}
\partial_+ z^{\underline{M}} B_{\underline{MN}} + ... \;,
\end{equation}
as well as the superstring kinetic term.
The constant $T$ is naturally interpreted as the
superstring tension.

To close this introductory section, we present the worldsheet gauge group
(including local $(8,0)$ supersymmetry) in the twistor formulation. The whole
action \p{twistact} and in particular $S_{wz}$ are invariant under the
following restricted class of diffeomorphisms of the $(8,0)$ superspace
\begin{eqnarray} \label{susy}
\delta \theta_\alpha &=& -{i\over 2} D_{\alpha} \Lambda^-\;, \nn \\
\delta x^- &=& \Lambda^- - {1\over 2} \theta_\alpha D_{\alpha} \Lambda^-\;,
\nn \\
\delta x^+ &=& \Lambda^+ \;,
\end{eqnarray}
with $\Lambda^{\pm} (z)$ being arbitrary unconstrained superfield
parameters. This group is chosen so that it leaves the ``almost flat"
derivatives
\begin{eqnarray} \label{covder}
D_{\alpha} &=& \partial_{\alpha} + i\theta_{\alpha} \partial_- +
e_\alpha^{\;\;+}\partial_+ \nn \\
D_- &=& \partial_- - {i\over 8} D_\alpha e_\alpha^{\;\;+} \partial_+ \nn
\\
D_+ &=& \partial_+ \;
\end{eqnarray}
covariant. In (\ref{covder}) the single non-trivial component of the
worldsheet superspace inverse
vielbeins (defined by $D_A = E_A^M\partial_M$) is $e^{\;\;+}_\alpha$. It is
needed to ensure covariance under
the shifts of $x^+$ and obeys the constraint ($\{ \}$ denotes the symmetric
traceless part)
\begin{equation} \label{cnstr}
D_{\{ \alpha} e_{\beta \}}^{\;\;+} = 0\;,
\end{equation}
so that the covariant derivatives $D_\alpha, D_-$ still form the flat algebra
\begin{equation} \label{susyalg}
\{ D_\alpha, D_\beta \} = 2i \delta_{\alpha\beta}D_-\;, \;\;\;
[D_\alpha, D_-] =
 0\;.
\end{equation}
The constraint (\ref{cnstr}) is produced by the term $S_2$ in the
superstring action
(\ref{twistact}).

{}From the explicit form of the covariant derivatives $D_\alpha, D_-,
D_+$ above one can read off the form of the worldsheet zweibein matrix.
In what follows we shall make use of the vielbeins
$E^{\pm}_M$ entering the WZ term (\ref{wz}):
\begin{eqnarray} \label{zweibein}
E^+_+ &=& 1\;,\;\; E^+_- = {i\over 8}\; D_\alpha e_\alpha^+\;,\;\;E^+_\mu =
- e^+_\mu + {1\over 8} \theta_\mu\; D_\alpha e^+_\alpha \;, \nn \\
E^-_+ &=& 0\;, \;\; E^-_- = 1\;, \;\; E^-_\mu = -i \theta_\mu \;.
\end{eqnarray}

Finally, we give the transformation laws of the covariant derivatives and
$e^+_\alpha$ under the restricted diffeomorphism group (\ref{susy})
\begin{eqnarray} \label{transder}
\delta D_\alpha &=& {i\over 2} (D_\alpha D_\beta)\Lambda^- D_\beta \nn \\
\delta D_- &=& -(D_-\Lambda^-)D_- + {i\over 2}(D_- D_\alpha \Lambda^-)
D_\alpha \nn \\
\delta D_+ &=& -\left( D_+\Lambda^+ + {i\over 2}(D_+D_\alpha \Lambda^-)
e_\alpha^{\;+} + {i\over 8} (D_+\Lambda^-) D_\alpha e_\alpha^{\;+} \right)
D_+ \nn \\
&& - (D_+\Lambda^-) D_- + {i\over 2}(D_+D_\alpha \Lambda^-)D_\alpha \nn \\
\delta e_\alpha^{\;+} &=& D_\alpha \Lambda^+ + {i\over 2} D_\alpha D_\beta
\Lambda^- e_\beta^{\;+}\;.
\end{eqnarray}

\vspace{0.4cm}
\noindent {\bf 3. Chiral fermion action.} Let us now turn to our
basic task. We shall identify the chiral
fermions with
the first components of some anticommuting $(8,0)$ superfields $\Psi^i (z)$.
The index $i$ is in general arbitrary, but for definiteness we shall
regard it as a vector $SO(n)$ index, for instance as the vector
index of $SO(32)$. The action which we propose for this superfield will
be the following addition to the superstring WZ term \p{wz}
\begin{equation} \label{fermact}
S_f = \int d^2x d^8 \theta
\left\{ P^{MN}
E^{\;+}_{[M}\; (i\; \partial_{N\}}
\Psi^i \Psi^i + {1\over 8} E^-_{N\}}
\nabla_\alpha \Psi^i \nabla_\alpha \Psi^i)
+
P^{\{\alpha \beta\}}
\nabla_\alpha \Psi^i \nabla_\beta \Psi^i \right\}\;.
\end{equation}
Here $P^{MN}$ is the same Lagrange multiplier as in the superstring action
(\ref{wz}) and $P^{\{\alpha \beta \}}$ is a new symmetric traceless
Lagrange multiplier, whose meaning will be clarified later on. The
covariant derivative $\nabla_\alpha$ is defined as follows:
\begin{equation} \label{nabla}
\nabla_\alpha = D_\alpha - (\partial_+ e^+_\alpha) \; w \;.
\end{equation}
The operator $w$ measures the weight
with respect to local dilatations with  the superfield parameter
$D_+\Lambda^+ + \ldots$ (see
the first term in the transformation law $\delta D_+$ in (\ref{transder})).
As follows from \p{transder}, the derivative
$D_+$ has weight $-1$, so the vielbein $E^{\;+}_M$ present in
\p{fermact} has weight $+1$. Thus, to achieve invariance of
\p{fermact} we should ascribe to the superfield
$\Psi^i$ the weight $-{1/2}$.
The derivative \p{nabla} is covariant when acting on a superfield
with an arbitrary weight $w$. It still satisfies the flat algebra
\p{susyalg} together with the modified covariant $x^-$ derivative (its
own weight is zero)
\begin{equation}
\nabla_- = D_- + {i\over 8}\; \partial_+ (D_\alpha e^+_\alpha) \; w \; .
\end{equation}
Note that the term with $\partial_N\Psi$ in \p{fermact} is covariant as it
stands, since the superfields $\Psi^i$ anticommute.

In the superstring action the internal symmetry
group of the chiral fermions, e.g., $SO(32)$ is gauged. Therefore one needs
to add the pull-back of the $D=10$ super-Yang-Mills connection
$E_\alpha^{\underline M} A_{\underline M}(\underline z)$ to the covariant
derivative (\ref{nabla}). However, this does not affect the algebra of the
derivatives, provided the connection $A_{\underline M}(\underline z)$
satisfies the
constraints of $D=10$ super-Yang-Mills  theory \cite{SYM}. Therefore, for
simplicity in what follows we shall omit the gauge field term in
$\nabla_\alpha$.

Now we are ready to explain why we propose the chiral
fermion superfield action just in the form \p{fermact}. First of all,
we wish to produce the standard component kinetic term of the
chiral fermions
$$
\sim \psi^i \partial_- \psi^i\;,\;\;\; \psi^i = \Psi^i|_{\theta =0}\;,
$$
using the same mechanism as in the case of the superstring WZ term, i.e.
by passing to the gauge \p{pgauge}. This accounts for the term
$\partial_N\Psi^i\Psi^i$ in \p{fermact}. It is of the same form as the one
proposed in \cite{ST}. However, the second $\Psi$ term in \p{fermact}
was not included in the action in \cite{ST}. The fact that it should be
there follows from two requirements:

(i) the integrability condition \p{hequat}, which is
equivalent to the $D=10$
supergravity constraints, should
not be affected by adding the action \p{fermact} to the original
superspace WZ term \p{wz};

(ii) the constraint on the superfields
$\Psi^i$ following from requirement (i) {\it should not put the theory
on shell}. Precisely this point accounts for the main difference between
our action and that of \cite{ST}.

Let us first explain how requirement (i) is satisfied. In the presence of
the new action term (\ref{fermact}) varying with respect to $P^{MN}$
leads to modifications of the two- and three-forms in eq. \p{bequat} and
its consequence \p{hequat}.
We denote them by $\hat B_{MN}$ and $\hat H_{KMN}$, respectively:
\begin{equation}
B_{MN} \Rightarrow  B_{MN} +  \hat B_{MN}\;,\;
H_{KMN} \Rightarrow H_{KMN } + \hat H_{KMN}\;,\;\;
\hat H_{KMN}= \partial_{[K } \hat B_{MN \}}\;.
\end{equation}
So, we require
\begin{equation} \label{hvanish}
\hat H_{[KMN \}}= 0 \;.
\end{equation}
To find the constraint on $\Psi$ following from (\ref{hvanish}), we replace
for the moment the expression $\nabla_\alpha \Psi^i \nabla_\alpha \Psi^i$
in \p{fermact} by an unknown $X$ and then write down
\begin{equation}
\hat B_{MN} =
E^{\;+}_M \;(i\; \partial_N
\Psi^i \Psi^i + {1\over 8}\; E^-_N\; X) - (-)^{MN}\; (M \leftrightarrow N)\;.
\end{equation}
Using the explicit form of the zweibein matrix \p{zweibein} it is
straightforward though a bit tedious to compute
all the components of $\hat H_{KMN}$ and to see that the
necessary and sufficient condition for them to vanish is
\begin{equation} \label{constr}
\nabla_\alpha \Psi^i \nabla_\beta \Psi^i - {1\over 8}\;
\delta_{\alpha \beta}\; X =
0 \;\Rightarrow X = \nabla_\alpha \Psi^i \nabla_\alpha \Psi^i \;.
\end{equation}
Let us give explicitly several components of $\hat H_{KMN}$
\begin{eqnarray}
\hat H_{\alpha \beta \gamma} &=& -2i \;E^+_{\alpha}\; [\;(
\nabla_\beta \Psi^i \nabla_\gamma \Psi^i  - {1\over 8}\;
\delta_{\beta \gamma}\; X)
\nn \\
&& - {1\over 16}\; \theta_\beta\; (
16i\; \nabla_- \Psi^i \nabla_\gamma \Psi^i - \nabla_\gamma X )\;] +
symmetrization  \label{1}\\
\hat H_{+\alpha \beta} &=& -2i \; (
\nabla_\alpha \Psi^i \nabla_\beta \Psi^i - {1\over 8}\;
\delta_{\alpha \beta}\; X) \nn \\
&&+ {i\over 8}\; \theta_\alpha\; (
16i\; \nabla_- \Psi^i \nabla_\beta \Psi^i - \nabla_\beta X) +
(\alpha \leftrightarrow \beta) \label{2} \\
\hat H_{\alpha + -} &=& {1\over 8}\; (\nabla_\alpha X - 16i\; \nabla_-\Psi^i
\nabla_\alpha \Psi^i)\;. \label{3}
\end{eqnarray}
{}From the vanishing of the last component it follows that
\begin{equation} \label{conseq}
16i\; \nabla_- \Psi^i \nabla_\beta \Psi^i - \nabla_\beta X = 0\;,
\end{equation}
then from the vanishing of the second one follows the constraint \p{constr}.
Eq. \p{conseq} is in fact a corollary of \p{constr}.

The most essential difference from the approach
of Sorokin and Tonin becomes clear at this point. They do not include
the $X$ term in the superstring WZ action and so from the same
consistency condition \p{hvanish} they derive the constraint
\begin{equation}\label{ston}
\nabla_\alpha \Psi^i \nabla_\beta \Psi^i = 0.
\end{equation}
It is much stronger than \p{constr} in that it implies the equations of
motion for $\Psi^i$. Below we will show that our constraint
\p{constr} is purely kinematical and serves to reduce the off-shell
field content of $\Psi^i$.

We would also like to comment on the approach of Howe \cite{Howe}. His
action is based on a {\it linear constraint} on the chiral fermion
superfield
$\Psi^{i' q}$:
\begin{equation}\label{howe}
D_\alpha \Psi^{i'q} = (\gamma^{i'})_{\alpha\dot\alpha} P^{\dot\alpha\; q}.
\end{equation}
Here the $SO(32)$ index is split into an $SO(8)$ index $i'$ and an $SO(4)$
index $q$. The indices $\alpha$ and $\dot\alpha$ are $s$ and $c$ spinor
indices of $SO(8)$ and $\gamma^{i'}$ is the $SO(8)$ gamma matrix. It
should be pointed out that eq. (\ref{howe}) provides a solution (but not
the general one) to our constraint (\ref{constr}) for this particular
arrangement of the indices. The advantage of the constraint (\ref{howe})
is its linearity, but its serious drawback (noted in \cite{Howe} as well)
is the strong restriction on the chiral fermion gauge group, which is at
most $SO(4)$ (or a tensor product of several $SO(4)$'s).

Coming back to our action \p{fermact}, it remains to explain the
r\^ole of the second Lagrange multiplier
term in it.  On the one hand, it serves
to impose the constraint (\ref{constr}) on shell. On the other hand, its
presence helps to maintain the gauge freedom \p{gauge1} which
allows one to bring $P^{MN}$ to the form \p{pgauge}.
The gauge variation of the $P^{MN}$ term in \p{fermact} is proportional to
the constraint \p{constr}, so it can be compensated by an appropriate
variation of the new Lagrange multiplier $P^{\{\alpha \beta\}}$.
The explicit form
of this variation is not too enlightening, so we do not give it here.
Note that $P^{\{\alpha \beta\}}$ has its own gauge freedom
\begin{equation} \label{gauge2}
\delta P^{\{\alpha \beta \}} = \nabla_\gamma \; \Lambda^{\{\gamma \alpha
\beta \}}\;,
\end{equation}
which allows one to gauge away some components of this superfield.

The meaning of the constraint \p{constr} and how it helps to make
the action \p{fermact} supersymmetric in the gauge \p{pgauge} will be
discussed in the next Section. In this gauge \p{fermact} takes the form
\begin{equation} \label{fermact2}
S_f =T \int d^2 x d^8\theta
\left\{\theta^8 \; (i\; \nabla_-
\Psi^i \Psi^i + {1\over 8}
\nabla_\alpha \Psi^i \nabla_\alpha \Psi^i) +
P^{\{\alpha \beta\}}
\nabla_\alpha \Psi^i \nabla_\beta \Psi^i \right\}\;.
\end{equation}

\vspace{0.4cm}
\noindent {\bf 4. Peculiarities of the chiral fermion action.}
Here we discuss some unusual features of the action \p{fermact2} as it
stands, leaving aside its superstring-inspired appearance.
It is worth mentioning that though we have
introduced it in the framework of the $D=10$ heterotic superstring and
of an $(8,0)$ worldsheet superspace, its basic features actually
do not depend on the Grassmann dimension of the worldsheet superspace. So we
may consider a general  $(N,0)$
superspace, which corresponds to the change $8\rightarrow N$ in
\p{fermact2}. We assume that the $\theta$'s are always real and
transform, in general, according to the vector representation of
the automorphism group
$SO(N)$ \footnote{In the case $N=8$ (as well as $N=4,2,1$), due to the
triality property of $SO(8)$,
the $\theta$'s can equally well be placed in either the $s$ or $c$ spinor
representations of $SO(8)$. This is in fact necessary if one wants to fix
a light-cone gauge in which the worldsheet $\theta$'s are identified with
one half of the target superspace ones (see \cite{DGHS} for details).}.
We also assume that the derivatives $\nabla_\alpha,
\nabla_-$ in \p{fermact2}
are flat, i.e. we put $e^+_\alpha = 0$. Actually, one can always choose a
gauge in which all covariant derivatives are almost flat \cite{DIS},
\cite{DGHS},
\begin{equation}
e^+_\alpha = i\theta_\alpha \;g_{--}\;, \;\;\; D_- = \partial_- +
g_{--}\;\partial_+\;,
\end{equation}
where $g_{--}$ is the component of the worldsheet metric responsible for
the second Virasoro constraint. For simplicity we put $g_{--}$
equal to zero (which is only possible locally), although this is not
essential for what follows.

First we demonstrate that the action
\begin{equation} \label{Nferm}
S = T\int d^2 x d^N\theta
\left\{ \theta^N \; (i\;
\partial_-
\Psi^i \Psi^i + {1\over N} \;
D_\alpha \Psi^i D_\alpha \Psi^i) +
P^{\{\alpha \beta\}}
D_\alpha \Psi^i D_\beta \Psi^i \right\}
\end{equation}
is supersymmetric despite the presence of explicit $\theta$'s.
The easiest way to see this is to observe that
\begin{equation}
i\;\partial_-
\Psi^i \Psi^i + {1\over N}
D_\alpha \Psi^i D_\alpha \Psi^i = {1\over N}\;D_\alpha (\Psi^i D_\alpha
\Psi^i)\;,
\end{equation}
then to integrate in both terms of the action by parts and, finally,
to rewrite it in a
very simple form resembling Chern-Simons type actions \cite{HT}
\begin{equation} \label{Nferm2}
S = T \int d^2x d^N \theta P^{\alpha}\;\Psi^i D_\alpha \Psi^i
\end{equation}
with
\begin{equation} \label{p}
P^{\alpha} = \theta^{N-1}_\alpha + D_\beta P^{\{\beta \alpha\}}\;,\;\;
\theta^{N-1}_\alpha  \equiv D_\alpha \theta^N\;
\end{equation}
(actually, when passing from \p{Nferm} to \p{Nferm2}, there appears a minus
sign in the case of an odd $N$, but it can be absorbed into the constant
$T$).

At this point we can forget the particular structure of $P^\alpha$ \p{p}
coming from the original definition of the chiral fermion action. Instead, we
can define the action by eq. \p{Nferm2} with $P^\alpha$ satisfying
the constraint
\begin{equation} \label{constrP}
D_\alpha P^\alpha = 0\;.
\end{equation}
It is easy to check that \p{p} provides the general solution to this
constraint up to an arbitrary right-moving function $a(x^+)$ in front of
the first (cohomological) term in (\ref{p}). Note, however,
that the action
(\ref{Nferm2}) and the constraint \p{constrP} are invariant under $x^+$
dependent scale transformations
\begin{equation}
\Psi^i \rightarrow \lambda^{1/2} (x^+) \Psi^i\;,\; P^{\alpha}
\rightarrow \lambda^{-1} (x^+) P^\alpha
\end{equation}
which allow us to gauge $a(x^+)$ into any non-zero constant. So,
the action \p{Nferm2} with the additional constraint \p{constrP}
describe the general situation.

In this Chern-Simons-like representation the chiral fermion action
does not include explicit $\theta$'s (they are hidden in the solution
of the constraint \p{constrP}) and is invariant (together with the
constraint) with respect to the local
transformations \p{susy} with $\Lambda^+ = \Lambda^+(x^+)$, provided
$P^\alpha$ transforms according to the law
\begin{equation} \label{Psusy}
\delta P^\alpha = -{i\over 2}\;(D_\beta D_\alpha \Lambda^-) P^\beta
+ {1\over 2} (N-2) (\partial_- \Lambda^-) P^\alpha \;.
\end{equation}
The second term in \p{Psusy} cancels the variation of
the superspace integration measure in \p{Nferm2}.

Now let us discuss some peculiar features of the realization
of rigid supersymmetry in the above action. To this end we
will go to components by varying in (\ref{Nferm}) with respect to the
Lagrange multiplier $P^{\{\alpha \beta \}}$, substituting the
constraint back into the action (this will be justified below)
and finally integrating over the $\theta$'s. The $\theta$ integration is
now trivial because of the presence of the factor
$\theta^N$. In this way we obtain that for an arbitrary $N$
the chiral fermions are described off shell by the action
\begin{equation} \label{fermcomp}
S_f = T\int d^2 x (i\;\partial_- \psi^i \psi^i + {1\over N}\;
b^i_\alpha b^i_\alpha )\;;\;\;\;
b^i_\alpha \equiv D_\alpha \Psi^i |_{\theta = 0}\;,
\end{equation}
supplemented by the nonlinear superfield constraint
\begin{equation} \label{constr2}
D_{\alpha} \Psi^i D_\beta \Psi^i - {1\over N}\;\delta_{\alpha \beta}\;
D_{\gamma} \Psi^i D_\gamma \Psi^i = 0\;.
\end{equation}

The first unusual feature of the action \p{fermcomp} is that, irrespective
of the value of $N$, it involves
only the two first fields from the $\theta$ expansion of $\Psi^i$.
Nevertheless, it is off-shell supersymmetric! To see this, we first
write down explicitly the first two component constraints following from the
superfield one \p{constr2}
\begin{eqnarray}
&& b^i_{\{\alpha} b^{i}_{\beta \}} = 0\;, \label{compconstr1} \\
&& \phi^i_{[\gamma \{\alpha]} b^i_{\beta\}} +
i \delta_{\gamma \{\alpha} \partial_- \psi^i b^i_{\beta\}} = 0
\label{compconstr2}
\end{eqnarray}
with
$$
\phi^i_{[\alpha \beta]} \equiv {1\over 2}\;D_{[\alpha} D_{\beta]} \Psi^i|_
{\theta = 0}\;.
$$
The $\gamma\beta$ trace of \p{compconstr2} yields the important
relation
\begin{equation} \label{constr3}
\phi^i_{[\alpha \beta]} b^i_\beta - i(N-1)\;\partial_-\psi^i b^i_\alpha
= 0 \;,
\end{equation}
while the part with mixed symmetry results in some further restriction
on $ \phi^i_{[\alpha \beta]} $ which does not involve the physical field
$\psi^i$ (its explicit form is not needed for our purposes).

Now, let us make a rigid supersymmetry transformation in \p{fermcomp} (the
transformation laws for the component fields follow from the expansion of
the superfield $\Psi^i$),
\begin{equation} \label{supsi}
\delta \psi^i = \epsilon_\alpha\;b^i_\alpha\;,\;\;\;\;
\delta b^i_\alpha = \epsilon_\gamma\;(\phi^i_{[\gamma \alpha]} + i\delta
_{\gamma \alpha}\;\partial_-\psi^i)\;.
\end{equation}
The variation of the action is
\begin{equation}
\delta S_f = T\int d^2x {2\over N} \epsilon_\gamma
[\phi^i_{[\gamma \alpha]}b^i_\alpha - i(N-1)\;\partial_-\psi^i b^i_\gamma
]
\end{equation}
and it vanishes in virtue of \p{constr3}. Thus, the action \p{fermcomp}
is supersymmetric due to the superfield constraint
\p{constr2}. As we have just seen, in fact only the component \p{constr3}
of this constraint is involved in achieving the supersymmetry
of the action; the other component constraints are needed to ensure
that the whole set of them are supersymmetric in their own right \footnote
{Note that for the supersymmetry of \p{fermcomp} it is enough to impose the
weaker constraint $D_\alpha D_{\{\alpha}\Psi^i D_{\beta\}}\Psi^i = 0$,
the first
component of which is just eq. \p{constr3}. The reason for choosing
\p{constr2} is the requirement
of consistency with the twistor-like formulation of the superstring,
as explained earlier. }.

Now we discuss the meaning of the component constraints.
Splitting $b^i_\alpha$ into a ``radial'' and an ``angular'' parts,
\begin{equation} \label{decompb}
b^i_\alpha = m \hat b^i_\alpha\;,
\end{equation}
we can rewrite \p{compconstr1} as
\begin{equation} \label{constrb}
m\;(\hat b^i_\alpha \hat b^i_\beta - {1\over N}\;\delta_{\alpha \beta}) =
0\;.
\end{equation}
If one assumes $m$ to be non-singular off-shell (this is not so
on-shell, see below), then \p{constrb} implies that $b^i_\alpha$ is
an orthogonal $SO(N)$ matrix if $n = N$ or it represents the coset
$SO(n)/SO(n-N)$ if $ n > N$.
In order for this to make sense we have to take $n \geq N$.
With the help of $\hat b^i_\alpha$ one can covariantly split any
$SO(n)$ vector into an $SO(n-N)$ projection and an
orthogonal $N$-dimensional one which is inert under $SO(n)$, but
is transformed in a proper way by the automorphism $SO(N)$. So
$\hat b^i_\alpha$ is a sort of a ``bridge'' relating these two groups.
The meaning of the constraint \p{constr3} becomes clear now: it
states that such an $SO(N)$ projection of the component
$\phi^i_{[\gamma \alpha]}$ for some irreducible combination of its
$SO(N)$ indices is not independent, but is expressed in terms of
$\partial_-\psi^i$. As was already mentioned,  \p{compconstr2}
implies more constraints on this projection which amount to
the vanishing of some other
of its $SO(N)$ irreducible components. The rest of
$\phi^i_{[\gamma \alpha]}$ remains arbitrary. For instance, in
the case $N=2$ we have only the constraint \p{constr3} on  the
$n$ fields $\phi^i = \epsilon_{\alpha \beta}\phi^i_{\alpha \beta}$, so
in the latter there remain $n-2$ independent components;
in the case $N=3$ we have five more
constraints of the second type alongside \p{constr3}, which together leave
$3n - 8$ independent
components in the
$3n$ fields $\phi_{[\alpha \beta]}$, etc. It can be shown that all of the
subsequent component constraints
in \p{constr2} have the typical structure of \p{compconstr2}:
they mean that some $b$ projections of the corresponding
higher components of $\Psi^i$ either vanish or are expressed in terms of the
lower ones.

It should be pointed out that a priori it is not so evident
that the constraint \p{constr2}
is purely kinematical and does not produce differential conditions of
the kind of equations of motion for some higher components of
$\Psi^i$ and for certain values of $N$. If it were the  case, we would not
be allowed to substitute the constraint back into the action. Fortunately,
simple arguments show that for any $N$ this constraint remains purely
kinematical off shell.

To convince ourselves, let us pull out a constant part from $\Psi_i$,
\begin{equation}
\Psi_i = \theta_\alpha \delta_{\alpha i} + \tilde \Psi_i
\end{equation}
and split the index $i$ into $i = ({\alpha}, {i'})$, where
${\alpha}$ and ${i'}$ run, respectively, from 1 to $N$ and
from $N+1$ to $n$. Then $\tilde \Psi_i$ splits into a pair of superfields,
\begin{equation}
\tilde \Psi_i = (\tilde \Psi_{{\alpha}}, \tilde \Psi_{{i'}}).
\end{equation}
In terms of these superfields the original constraint (\ref{constr2})
can be rewritten as a linear one,
\begin{equation} \label{linconstr2}
D_{\alpha} \Phi_{\beta} + D_{\beta} \Phi_{\alpha} -
\delta_{\alpha \beta} {2\over N} D_{\gamma} \Phi_{\gamma} = 0,
\end{equation}
where
\begin{equation} \label{rel}
\Phi_{\alpha} \equiv \tilde \Psi_{\alpha} +
{1\over 2} \tilde \Psi_{{\beta}} D_{\alpha} \tilde \Psi_{{\beta}}
+ {1\over 2} \tilde \Psi_{{i'}} D_{\alpha} \tilde \Psi_{{i'}}\; .
\end{equation}
Thus, we have a purely
kinematical constraint for the superfield $\Phi_{\beta}$ which
has a simple general solution:
\begin{equation}
\Phi_\alpha = D_\alpha G (z)  + \theta_\alpha F(x^+, x^-)\;.
\end{equation}
On the other hand, the relation
(\ref{rel}) is just a canonical redefinition of $\tilde
\Psi_{\alpha}$: one can reexpress it in terms of $\Phi_{\alpha}$ and
$\tilde \Psi_{{i'}}$  from (\ref{rel}) by means of
iterations. The superfield $\tilde \Psi_{{i'}}$
remains entirely
unconstrained. So, it becomes clear that for any $N$
the constraint (\ref{constr2}) does not contain any dynamics off shell
and merely expresses some components of $\Psi_i$ in terms of others
(or puts them equal to zero). The corresponding Lagrange multiplier
$P^{\{\alpha \beta\}}$ does not contain propagating degrees
of freedom. We note that just because the
constraint \p{constr2} is kinematical, there is no local
symmetry of Siegel's type \cite{Siegel} associated with
$P^{\{\alpha \beta\}}$ in the actions \p{fermact}, \p{fermact2},
\p{Nferm}, \p{Nferm2}. Once again, this is in contrast to ref. \cite{ST},
where the constraint (\ref{ston}) is dynamical and its Lagrange multiplier
needs some sort of
Siegel's invariance in order not to propagate.

Thus we have seen that there are no problems with the off-shell supersymmetry
of the chiral fermion action. However, going on shell in this
action is rather subtle. As follows from \p{fermcomp}, on shell
\begin{equation} \label{eqmob}
b^i_\alpha = 0
\end{equation}
and we are no longer allowed to divide by $m$ in the constraints
\p{compconstr1}, \p{compconstr2} and the subsequent ones. At first
sight this does not lead to difficulties as
\p{compconstr1}, \p{compconstr2} are satisfied identically if $b^i_\alpha$
satisfies \p{eqmob}. However, beginning with $N=3$, in the higher-order
constraints there appear terms which are not multiplied by $b^i_\alpha$
and so do not vanish on shell. For instance, in the $N=3$ case on shell
there remains the following quadratic constraint on
the field $\phi^i_{[\gamma \alpha]}$
\begin{equation} \label{badconstr}
\phi^i_{[\gamma \alpha]} \phi^i_{[\gamma \beta]}
\epsilon_{\alpha\beta\sigma} = 0\;.
\end{equation}
For higher $N$ there appear quadratic constraints of this type including
derivatives of fields. This would seem to create difficulties if these fields
were somehow present in the action. However, the only auxiliary field
entering the component action (\ref{fermcomp}) is in fact
the ``radial part" $m$ of $b^i_\alpha$. The others are needed only to
supersymmetrize the constraint \p{constr3} (itself required for the
supersymmetry of the action). This indicates that the appearance of
such strange constraints on shell is harmless. When we vary with respect
to the fields $\psi^i$ and $m$ in the action  to obtain the
equations of motion, we should vary the constraints as well. The latter
simply serve to partially express the variations of the
higher-order components of $\Psi^i$ through $\delta \psi^i$ and $\delta m$.
The presence of the terms of the type \p{badconstr} in the component
constraints results in  factors of $m^{-1}$ in the expressions for the
variations of the higher-order fields.  The more we approach
the minimum, the bigger the variations of these fields become, tending to
infinity at the minimum itself. But we should not care about these
variations as they do not contribute to the action functional,
i.e. do not generate any equations of motion.  Another argument is that on
shell $(N,0)$ supersymmetry is trivially realized. Indeed, there the
supersymmetry variation (\ref{supsi}) of $\psi^i$ vanishes, but this does
not contradict the algebra of supersymmetry as on shell
\begin{equation}
\partial_-\psi^i = 0.
\end{equation}

In conclusion we can say that the form of the chiral fermion action with
manifest (local) worldsheet supersymmetry presented here completes the
twistor-like formulation of the heterotic string in $D=10$ and can
serve as a starting point for a new attempt to covariantly quantize
the theory.
\vskip5mm

\noindent {\bf Acknowledgements} The authors have profited from discussions
with F. Delduc, A. Galperin and V. Fateev. E.I. would like to thank Prof.
V. Rittenberg for hospitality at the Physics Institute of the University of
Bonn, where this work has been done. E.I. is also grateful to the Russian
Foundation of Fundamental Research, grant 93-02-3821, and to the
International Science Foundation, grant M9T000, for financial support.


\begin{thebibliography}{22}

\bibitem{kniga}M. B. Green, J. H. Schwarz and E. Witten, {\it Superstring
Theory}, CUP, 1987.
\bibitem{Ton}M. Tonin, {\sl Phys. Lett.} {\bf 283B} (1992) 213;
{\sl Int. J. Mod. Phys.} {\bf A7} (1992) 6013.
\bibitem{DGHS} F. Delduc, A. Galperin, P. Howe and E. Sokatchev, {\sl Phys.
Rev.} {\bf D47} (1992) 578.
\bibitem{gr}S. Aoyama, P. Pasti and M. Tonin, Phys. Lett. {\bf 283B} (1992)
213;\\ I. Bandos, D. Sorokin, M. Tonin and D. Volkov, {\it Phys. Lett.}
{\bf 319B} (1993) 445.
\bibitem{nonhet} A. Galperin and E. Sokatchev, {\sl Phys.
Rev.} {\bf D48} (1993) 4810; \\
V. Chikalov and A. Pashnev, {\it Mod. Phys. Lett.} {\bf A8} (1993)
285; \\
P. Pasti and M. Tonin, preprint DFPD/94/TH/05, Padova (1994);\\
I. Bandos, M. Cederwall, D. Sorokin and D. Volkov, preprint
ITP-94-10, G\"oteborg (1994).
\bibitem{p-b} P. Pasti and M. Tonin, {\sl Nucl. Phys.} {\bf B418}
(1994) 337; \\
E. Bergshoeff and E. Sezgin, preprint CTP TAMO-67/93.
\bibitem{volk}D. P. Sorokin, V. I. Tkach and
D. V. Volkov, {\sl Mod. Phys. Lett.} {\bf A4} (1989) 901; \\
D. P. Sorokin, V. I. Tkach, D. V. Volkov and A. A. Zheltukhin,
{\sl Phys. Lett.} {\bf 216B} (1989) 302.
\bibitem{DS}F. Delduc and E. Sokatchev, {\sl Class. Quantum  Grav.} {\bf
9} (1991) 361.
\bibitem{GS}A. Galperin and E. Sokatchev, {\sl Phys. Rev.} {\bf
D46} (1992) 714.
\bibitem{IK}E. A. Ivanov
and A. A. Kapustnikov, {\sl Phys. Lett.} {\bf 267B} (1991) 175.
\bibitem{berko}N. Berkovits, {\sl Phys. Lett.} {\bf
232B} (1989)184; {\bf 241B} (1990) 497; {\sl Nucl. Phys.} {\bf B350}
(1991) 193; {\bf B358} (1991) 169.
\bibitem{DIS}F. Delduc,
E. Ivanov and E. Sokatchev, {\sl Nucl. Phys.} {\bf B384} (1992) 334.
\bibitem{ST}D. Sorokin and M. Tonin, {\sl Phys. Lett.} {\bf 326B} (1994) 84.
\bibitem{Howe}P. Howe, {A note on chiral fermions and heterotic strings},
King's College preprint, 1994.
\bibitem{10sg}B. E. W. Nilsson, {\sl Nucl. Phys.} {\bf B188} (1981) 176.
\bibitem{SYM}E. Witten, {\sl Nucl. Phys.} {\bf B266} (1986) 245.
\bibitem{HT}P. Howe and P. Townsend, Phys. Lett. {\bf 259B} (1991) 285.
\bibitem{Siegel} W. Siegel, {\sl Nucl. Phys.} {\bf B238} (1984) 307.


\end{thebibliography}
\end{document}